\documentclass{iopart}

\usepackage{iopams,array,enumerate,graphicx}

\begin{document}

\title[The instability of downside risk measures]{The instability of downside risk measures}
\author{I Varga-Haszonits$^{1,2}$ and I Kondor$^{1,3,4}$}
\address{$^1$ Department of Physics of Complex Systems, E\"otv\"os University, P\'azm\'any P\'eter s\'et\'any 1/A, H-1117 Budapest, Hungary}
\address{$^2$ Analytics Department of Fixed Income Division, Morgan Stanley Hungary Analytics, De\'ak Ferenc u. 15, H-1052 Budapest, Hungary}
\address{$^3$ Collegium Budapest -- Institute for Advanced Study, Szenth\'aroms\'ag u. 2, H-1014 Budapest, Hungary}
\address{$^4$ Parmenides Center for the Study of Thinking, Kardinal Faulhaber Strasse 14a, Munich, D-80333, Germany}
\eads{\mailto{Istvan.Varga-Haszonits@morganstanley.com}, \mailto{kondor@colbud.hu}}
\begin{abstract}
We study the feasibility and noise sensitivity of portfolio optimization under some downside risk measures (Value-at-Risk, Expected Shortfall, and semivariance) when they are estimated by fitting a parametric distribution on a finite sample of asset returns. We find that the existence of the optimum is a probabilistic issue, depending on the particular random sample, in all three cases. At a critical combination of the parameters of these problems we find an algorithmic phase transition, separating the phase where the optimization is feasible from the one where it is not. This transition is similar to the one discovered earlier for Expected Shortfall based on historical time series. We employ the replica method to compute the phase diagram, as well as to obtain the critical exponent of the estimation error that diverges at the critical point. The analytical results are corroborated by Monte Carlo simulations. 

\noindent{\it Keywords\/}: Replica Method, Critical Phenomena, Portfolio Optimization, Expected Shortfall, Value-at-Risk, Semivariance
%Critical phenomena of socio-economic systems, Risk measure and management, Cavity and replica method
\end{abstract}
\pacs{89.65.Gh}
\submitto{JSTAT}
\maketitle

\section{Introduction}

Portfolio optimization is one of the fundamental problems of financial theory. The first treatment of the topic appeared in the famous work by Markowitz \cite{markowitz}, who measured risk by the standard deviation of asset price fluctuations. In this context, portfolio optimization consists in minimizing the variance of the portfolio return given the expected return and the budget constraint. Although this defines a straightforward mathematical problem, the statistical properties of the solution turn out to be non-trivial when the covariances of the asset returns are estimated from a finite sample.

An extensive investigation of the noise sensitivity of the Markowitz portfolio optimization problem \cite{noisy1,noisy2, noisy3} revealed that for normally distributed asset returns the expected value of the ratio $q_0$ of the risk of the estimated optimum and that of the true optimum is proportional to $(1-N/T)^{-1/2}$, where $N$ is the number of assets in the portfolio and $T$ is the sample size (number of observation periods). In other words, the estimation error diverges as $T\to N$, and, in order to reduce the estimation error to a reasonable level, one needs a fairly large sample. Moreover, the estimated optimal portfolio weights exhibit dramatic fluctuations from one sample to another, and these fluctuations decay very slowly with increasing sample size. Covariance matrix filtering techniques based on Bayesian Shrinkage \cite{ledoit1,ledoit2,ledoit3} and Random Matrix Theory \cite{laloux1,laloux2,plerou1,plerou2,burda2} were shown to effectively reduce $q_0$ \cite{noisy2}, however, these techniques do not generally suppress the large fluctuations of the estimated portfolio weights.

In addition to the noise sensitivity of the classical standard deviation, Kondor et al \cite{kondor1} also examined the sensitivity of portfolio optimization under a few alternative risk measures, such as Mean Absolute Deviation \cite{konno}, Maximal Loss \cite{young} and Expected Shortfall \cite{acerbi1,acerbi2}. All of these were found to be even more susceptible to sample fluctuations than standard deviation, and in addition, Expected Shortfall (and Maximal Loss as its special case) displayed an additional instability in that the very existence of the optimum turned out to depend on the sample and the probability of the existence of an optimum was found to be less than one for any finite sample size. In other words, even if Expected Shortfall has a well defined minimum for a given asset return distribution, it may not have an optimum on a finite sample generated by that distribution.

Expected Shortfall is perhaps the simplest and intuitively most appealing example of the celebrated Coherent Risk Measures \cite{artzner1,artzner2}, which were introduced in response to the widespread use of ad-hoc risk measures (including Value-at-Risk) with poor theoretical foundation and well-known shortcomings. However, the instability discovered on the example of Expected Shortfall raised the suspicion that Coherent Risk Measures, all their axiomatic beauty notwithstanding, may be highly susceptible to sampling error in general. Indeed, this conjecture has been proved to be true by showing that no Coherent Measure of Risk has a minimum, if there exists a portfolio that produces positive returns for all observations on the given sample \cite{kondorvhi}.

The studies mentioned above were based on non-parametric estimators of the risk measures in consideration, without any a priori assumption about the sample generating process. However, estimators based on historical time series are notoriously unstable, so it is legitimate to ask whether parametric estimation could suppress the instability. Moreover, Value-at-Risk is often measured in practice by parametric estimation using some assumption about the probability distribution of the asset returns \cite{jorion}. Since in practice VaR is the most important measure in use today, and, furthermore, its parametric estimation is analogous to that of ES, it is a natural idea to study the stability of portfolio optimization under VaR and ES by fitting a multivariate Gaussian distribution on the sample of asset returns. The main objective of this paper is to decide whether the instability of historical ES (and VaR) estimation can be circumvented by parametric estimation. For the sake of simplicity we are also going to assume that the data generating process is itself Gaussian. It will turn out that, although parametric fitting reduces the instability, it does not eliminate it.

It should be noted that this paper, as well as the earlier studies mentioned above, investigate the noise sensitivity of the global risk minimization, without imposing any constraint on the expected return. This is a special case of the practically more relevant risk-reward optimization problem. It is clear, however, that adding a linear constraint to the global minimum risk problem does not change essentially the noise sensitivity characteristics. Focusing on the simpler problem makes it easier to understand and identify the effects and consequences of sampling error, while at the same time leaves open the possibility of revisiting the more general problem later.

The rest of the paper is organized as follows. \Sref{sec:esnonparam} is a brief overview of earlier results on the instability of the minimization of Expected Shortfall with non-parametric estimation. In \Sref{sec:esvarmin} we solve the ES/VaR minimization problem assuming that asset returns follow a multivariate normal distribution with explicitly known means, variances and covariances, and we derive the condition for the solution to exist. In \Sref{sec:esvarnoise} we investigate the feasibility and noise sensitivity of ES/VaR minimization when the parameters of the asset return distribution are estimated from finite samples. This section, which constitutes the backbone of our paper, is divided into several subsections: in \ref{sec:charnoise} we introduce some notations and terminology, in \ref{sec:replica} we use the replica method to characterize the critical behavior of the finite sample instability of the optimization problem, in \ref{sec:corr} we generalize these results to the case of correlated asset returns, in \ref{sec:montecarlo} we back up our findings with simulation, and finally in \ref{sec:semivar} we apply our results to the special case of semivariance minimization. The paper ends on a brief summary.

\section{The noise-sensitivity of Expected Shortfall minimization with non-parametric estimation}
\label{sec:esnonparam}

To put our discussion in context, we provide a brief overview of the results for the minimization of Expected Shortfall using a non-parametric estimator. Expected Shortfall is the mean value of losses exceeding a high threshold (referred to as the confidence level) specified in probability rather than in money. For instance, at confidence level $\alpha$ the Expected Shortfall ($\mbox{ES}_{\alpha}$) of an investment is the average of losses that occur in the $(1-\alpha)100$ percent of the worst cases. 

Historical ES based on a finite sample consisting of $T$ observations can be estimated by sorting these observations into ascending order and computing the average of the $T(1-\alpha)$ smallest values. Special care must be taken, however, when $T(1-\alpha)$ is not an integer number: in such a case one of the observations has to be 'split'. (For the precise definition of $\mbox{ES}_{\alpha}$ see for instance \cite{acerbi2}.) It was shown in \cite{rockafellar} that within this scheme portfolio optimization is equivalent to a convex linear programming problem. This is to be contrasted with the case of VaR, which, as a quantile, has no reason to be convex, and, indeed, is often found to be non-convex when estimated from historical time series. (This is why the problem of the noise sensitivity of VaR was ignored in \cite{kondor1}: in a sense \textit{historical} VaR is always unstable.) The highly desirable property of convexity has made ES very popular with academics, though ES is still very far from replacing VaR in practice or regulation.

\begin{figure}
\begin{center}
\includegraphics[keepaspectratio,width=10cm]{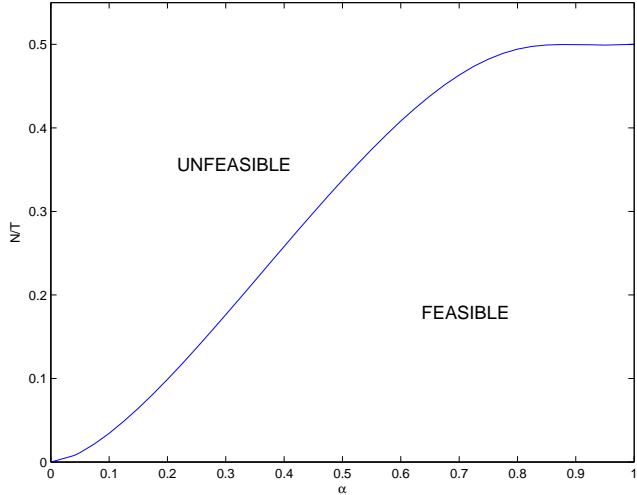}
\caption{\label{fig:esnonparam}The boundary between the feasible and unfeasible phases of the Expected Shortfall minimization problem on the $N/T$ vs $\alpha$ plane, in the $N\to\infty$ and $N/T=\mbox{const}$ limit.}
\end{center}
\end{figure}

As mentioned in the introduction, the noise sensitivity of ES optimization was examined in \cite{kondor1}. That study used a simulation based approach and assumed, for simplicity, iid normal asset returns. The (linear programming based) portfolio optimization algorithm was performed on a large number of such samples and the existence and distribution of the solution was investigated. The main findings of this study are the following:
\begin{itemize}
\item ES as a risk measure is much more sensitive to sample to sample fluctuations than the variance.
\item On some samples ES does not even have a minimum but diverges to minus infinity.
\item The probability of the existence of the optimum depends on the confidence level $\alpha$, as well as on the ratio between the number of assets $N$ and the number of observations $T$.
\item In the limit where $N\to\infty$ and $N/T$ is held constant the probability of the existence of the optimum tends either to 1 or to 0. On the $N/T$ vs $\alpha$ plane the zero probability (unfeasible) and unit probability (feasible) regions are separated by a well defined curve (the phase diagram), which was first determined by simulations \cite{kondor1}, then computed analytically by the replica method \cite{ciliberti1} (see Figure \ref{fig:esnonparam}).
\end{itemize}
In practical applications the confidence level is typically $\alpha>0.9$, and as shown by Figure \ref{fig:esnonparam} in that region the critical $N/T$ ratio is very close to $1/2$. This means that in the practically relevant cases one must have at least twice as many observations as the number of assets in order to ensure even the mere existence of an optimal portfolio. (And, of course, a much larger sample is needed to make the estimation error reasonably low.) Moreover, the critical value of the ratio $N/T$ decreases for decreasing confidence level, which implies that ES optimization becomes more and more unstable, requiring larger and larger samples to give a meaningful result.

\section{The minimization of parametric ES and VaR for Gaussian asset returns}
\label{sec:esvarmin}

The sensitivity of Expected Shortfall to sample fluctuations casts a shadow of doubt on its practical applicability in portfolio selection. However, one may wonder  whether this instability is not due to the use of raw data in historical estimation and whether a parametric method might be more robust against sample to sample fluctuations.\footnote{We are obliged to M. Gordy for a stimulating discussion on this point.}  In order to decide the question, we are going to look into the noise sensitivity of portfolio selection in the simplest setting, that is when the underlying process is iid normal and when the risk is estimated by fitting a normal distribution to the sample. This is a standard procedure for VaR estimation \cite{jorion}, but the ES and VaR estimators are so closely related that we can examine them together.

When the return of an asset $X$ is normally distributed with mean $\mu$ and standard deviation $\sigma$, then both its VaR and ES can be written in the form
\begin{equation}
\mathcal{R}(X)=\phi(\alpha)\sigma-\mu.\label{eq:paramrisk}
\end{equation}
The particular form of the function $\phi(\alpha)$ depends on whether we are computing VaR or ES:
\begin{equation}
\phi(\alpha)=\cases{
	\Phi^{-1}(\alpha) & \mbox{for VaR,}\\
	-\frac{1}{1-\alpha}\int_0^{1-\alpha}\Phi^{-1}(p)dp=\frac{e^{-\frac{1}{2}[\Phi^{-1}(\alpha)]^2}}{(1-\alpha)\sqrt{2\pi}} & \mbox{for ES.}\\
	}\label{eq:phivares}
\end{equation}
where $\Phi^{-1}(x)$ denotes the inverse of the standard normal cumulative distribution function (or error function):
\begin{equation}
\Phi(x)=\frac{1}{\sqrt{2\pi}}\int_{-\infty}^{x}e^{-\frac{y^2}{2}}dy.
\end{equation}
We assume that $\phi(\alpha)$ is nonnegative and invertible in its domain\footnote{This means that for VaR we only allow confidence levels between $0.5$ and $1$. This is, however, not a real restriction, since VaR does not make sense as a risk measure for $\alpha<0.5$.}, and we will often omit its dependence on $\alpha$ in the notation. All the relevant quantities depend on $\alpha$ only through $\phi(\alpha)$.

Let us now assume that we have $N$ assets in the portfolio and their returns $x_i$ follow a multivariate normal distribution with means $\mu_i$ and variances/covariances $\sigma_{ij}$ (where $i,j=1,2,...,N$). A portfolio is simply a vector with components $w_i$ representing the amount invested in asset $i$. Then the expected value and the variance of the portfolio return will be $\sum_{i=1}^Nw_i\mu_i$ and $\sum_{i,j=1}^N\sigma_{ij}w_iw_j$, respectively. According to \eref{eq:paramrisk},  ES and VaR can then be written as:
\begin{equation}
\mathcal{R}_{\phi}(\{w_i\})=\phi\sqrt{\sum_{i=1}^N\sum_{j=1}^N\sigma_{ij}w_iw_j}-\sum_{i=1}^N\mu_iw_i.\label{eq:pfrisk}
\end{equation}
The optimal portfolio can be found by minimizing $\mathcal{R}(\{w_i\})$ subject to the budget constraint
\begin{equation}
\sum_{i=1}^Nw_i=1.
\end{equation}
It is easy to see that this optimization problem is equivalent to minimizing the following Lagrangean:
\begin{equation}\fl
\mathcal{L}(\{w_i\},z,\lambda,\eta)=\phi\sqrt{z}-\sum_iw_i\mu_i+\lambda\left(\sum_iw_i-1\right)+\eta\left(\sum_{ij}w_i\sigma_{ij}w_j-z\right).\label{eq:lagrange}
\end{equation}
where $\lambda$ is used to enforce the budget constraint while $z$ and $\eta$ have been introduced to make the objective function quadratic in the portfolio weights. The minimization of $\mathcal{L}$ is a routine task, and it turns out that the optimum exists if and only if the covariance matrix $\sigma_{ij}$ is non-singular and
\begin{equation}
B^2-AC+A\phi^2>0,\label{eq:condexist}
\end{equation}
where we introduced the notations $A=\sum_{ij}\sigma^{-1}_{ij}$, $B=\sum_{ij}\sigma^{-1}_{ij}\mu_j$ and $C=\sum_{ij}\mu_i\sigma^{-1}_{ij}\mu_j$. As long as these conditions are satisfied, the solution is given by
\begin{eqnarray}
w_i^*&=\frac{1}{2\eta^*}\sum_j\sigma^{-1}_{ij}(\mu_j-\lambda^*),\\
\lambda^*&=\frac{B}{A}-\left[\left(\frac{B}{A}\right)^2-\frac{C-\phi^2}{A}\right]^{1/2},\\
\eta^*&=\frac{1}{2}\left[\left(\frac{B}{A}\right)^2-\frac{C-\phi^2}{A}\right]^{1/2}.
\end{eqnarray}

Condition \eref{eq:condexist} makes it clear that the existence of an optimal portfolio is not automatically guaranteed, but depends on the parameters of the underlying distribution (specifically on the expected values and covariances of the asset returns). When these parameters are estimated from a random sample, the fulfillment or violation of \eref{eq:condexist} (i.e. the feasibility of the optimization problem) will also be a random event.

\section{The stability of parametric ES and VaR optimization on finite samples}
\label{sec:esvarnoise}

\subsection{The characterization of noise sensitivity}
\label{sec:charnoise}

Let us now assume the position of an investor who knows that the returns are Gaussian, but does not know the parameters (i.e. the means, variances and covariances) of the distribution, so she has to estimate them from a finite sample. Let us assume she makes $T$ independent observations, each consisting of a vector of $N$ realized asset returns. This sample can be represented by an $N\times T$ matrix with elements $x_{it}$ equal to the realized return of asset $i$ over time period $t$ ($i=1,2,...,N$ and $t=1,2,...,T$). The means $\mu_i$ and covariances $\sigma_{ij}$ can be estimated by the unbiased estimators
\begin{eqnarray}
\hat{\mu}_i&=\frac{1}{T}\sum_{t=1}^Tx_{it},\label{eq:meanest}\\
\hat{\sigma}_{ij}&=\frac{1}{T-1}\sum_{t=1}^T\left(x_{it}-\hat{\mu}_i\right)\left(x_{jt}-\hat{\mu}_j\right).\label{eq:covest}
\end{eqnarray}
Then the risk of portfolio $\{w_i\}$ can be estimated by substituting $\hat{\mu}_i$ and $\hat{\sigma}_{ij}$ into \eref{eq:pfrisk}. Let us denote this estimated risk by $\hat{\mathcal{R}}_{\phi}(\{w_i\})$. Now we can ask two fundamental questions:
\begin{enumerate}
\item\label{item:doesexist} Does $\hat{\mathcal{R}}_{\phi}(\{w_i\})$ have a minimum?
\item\label{item:howfar} If it does, how far is this minimum from the real optimum?
\end{enumerate}
Question \eref{item:doesexist} can be answered by checking whether condition \eref{eq:condexist} is fulfilled by $\hat{\mu}_i$ and $\hat{\sigma}_{ij}$. Moreover, the matrix $\hat{\sigma}_{ij}$ is positive semidefinite by construction, therefore the estimated optimum is unique, provided that it exits. As for Question \eref{item:howfar}, first we need to specify how to measure the distance from the real optimum. To this end, we use the generalization of the measure $q_0$ introduced in \cite{noisy2}, which in the present case is defined as follows.

If we know the parameters  $\mu_i$ and $\sigma_{ij}$ of the data generating distribution -- for instance, in a simulation study like the ones in \cite{noisy2} and \cite{kondor1} -- we explicitly know the true risk function $\mathcal{R}_{\phi}(\{w_i\})$. Let us assume that the data generating process is such that $\mathcal{R}_{\phi}$ has a minimum under the budget constraint, and let us denote the corresponding optimal weights by $w_i^*$. Our hypotetical investor, however, only knows the estimators $\hat{\mu}_i$ and $\hat{\sigma}_{ij}$, so she will minimize the estimated risk function $\hat{\mathcal{R}}_{\phi}(\{w_i\})$. Assuming that it exists, let this estimated optimum be $\hat{w}_i^*$. Although the investor might have the impression that this portfolio has risk $\hat{\mathcal{R}}_{\phi}(\{\hat{w}_i^*\})$ we know that its real risk is $\mathcal{R}_{\phi}(\{\hat{w}_i^*\})$ which, by definition, is greater than the risk in the true optimum $\mathcal{R}_{\phi}(\{w_i^*\})$. Therefore, the quantity
\begin{equation}
q_0=\frac{\mathcal{R}_{\phi}(\{\hat{w}_i^*\})}{\mathcal{R}_{\phi}(\{w_i^*\})}
\end{equation}
is a natural dimensionless measure of the distance of the estimated optimum from the true optimum. Moreover, the number $q_0-1$ has a straightforward interpretation: it is the percentage increase in the optimal risk the investor has to face due to the sampling error.

The properties of $q_0$ have been extensively studied both numerically and analytically for the case of global variance optimization \cite{noisy2,noisy3,burda1}. Let us briefly summarize the main findings of these investigations:
\begin{itemize}
\item $q_0$ is a random variable which fluctuates from sample to sample, and its distribution depends on $N$ and $T$.
\item For large $N$ and $T$ and their ratio kept constant, $\mathbb{E}q_0^2=(1-N/T)^{-1}$   ($\mathbb{E}$ denotes the average over sample fluctuations).
\item In the same limit ($N/T$ is held constant and $N\to\infty$) the variance of $q_0^2$ vanishes.
\end{itemize}
In other words, the estimation error $q_0$ is a self-averaging quantity, and for large $N$ and $T$ its average only depends on the ratio $r=N/T$. The divergence of $q_0$ in the limit $r\to1$ can be regarded as the manifestation of an algorithmic phase transition, with a critical point $r_c=1$ and a critical exponent $-1/2$ for the estimation error $q_0\sim(r_c-r)^{-1/2}$.

Further studies of the noise sensitivity of portfolio optimization led to the conclusion that the critical behavior of the estimation error is similar to the above for a number of other risk measures (e.g. mean absolute deviation, maximal loss, non-parametric Expected Shortfall \cite{kondor1}) and data generating processes (e.g. GARCH \cite{vhi}). As we shall see in the following section, parametric ES and VaR also belong to the same universality class.

\subsection{The replica approach}
\label{sec:replica}

Averaging over samples is the same as what is called quenched averaging (see e.g. \cite{mezparvir}) in the statistical physics of disordered systems. Therefore, the heuristic replica method that has been so successful in that field can also be used effectively to investigate the noise sensitivity of portfolio optimization \cite{ciliberti1,ciliberti2}. In this section, we are going to employ the replica approach 1) to determine under what circumstances the optimum exist, and 2) to compute $q_0^2$ provided that there is an optimum. The computations will be performed in the 'thermodynamic' limit, that is when $N\to\infty$ while $r=N/T$ is finite and fixed.

For the sake of simplicity we are going to assume that the data generating distribution is iid standard normal, in other words, the elements $x_{it}$ of the sample matrix are identically distributed and mutually independent standard normal random variables. (We use these assumptions to make our argument more straightforward, but as we shall see in the next subsection, introducing correlations into the model does not affect our main results.) Since in this case $\mu_i=0$ and $\sigma_{ij}=\delta_{ij}$, the true risk of a portfolio $\{w_i\}$ will be
\begin{equation}
\mathcal{R}_{\phi}(\{w_i\})=\phi\sqrt{\sum_{i=1}^Nw_i^2}.\label{eq:truerisk}
\end{equation}
For later convenience, we are going to use a modified form of the budget constraint:
\begin{equation}
\sum_{i=1}^N w_i = N,\label{eq:budconst}
\end{equation}
which obviously does not change the nature of the optimization problem (it only rescales the result by a factor of $N$). Thus, the minimum of \eref{eq:truerisk} subject to \eref{eq:budconst} will be the portfolio with weights $w^*_1=w^*_2=...=w^*_N=1$, and the minimal risk will be $\mathcal{R}_{\phi}^*=\phi\sqrt{N}$. Hence, for a standard normal data generating distribution the distance of a portfolio $\{w_i\}$ from the true optimum is given by
\begin{equation}
q_0^2=\frac{1}{N}\sum_{i=1}^Nw_i^2.\label{eq:q02def}
\end{equation}
(It is worth noting that in the special case of iid standard normal returns, we get exactly the same formula, if we measure the risk by standard deviation.)

It is clear that VaR/ES optimization based on a sample $\{x_{it}\}$ can be regarded as a statistical physics problem. Combining equations \eref{eq:lagrange}, \eref{eq:meanest} and \eref{eq:covest} the Hamiltonian of the problem can be written as
\begin{eqnarray}\fl
\mathcal{H}\left(\{w_i\},z,\eta;\{x_{it}\}\right)=N\phi\sqrt{z}-\frac{N}{T}\sum_{i=1}^N\sum_{t=1}^Tw_ix_{it}+\nonumber\\
+\eta\left[\sum_{t=1}^T\left(\sum_{i=1}^N\left(x_{it}-\frac{1}{T}\sum_{s=1}^Tx_{is}\right)w_i\right)^2-Tz\right],\label{eq:hamilt}
\end{eqnarray}
where we replaced the factor $1/(T-1)$ by $1/T$ in equation \eref{eq:covest}, which makes no difference in the thermodynamic limit. (The budget constraint is not explicitly included in the Hamiltonian, but it will be taken into account soon.) We are interested in finding the ground state of this system. It is expedient, however, first to introduce a fictitious inverse temperature $\beta$ and work out the partition function $Z$ for finite temperature. The partition function is a functional of $\phi(\alpha)$ and the sample $x_{it}$: 
\begin{eqnarray}\fl
Z_{\beta}\left[\phi;\{x_{it}\}\right]=\int_{-\infty}^{\infty}\prod_{i=1}^Ndw_i\int_{0}^{\infty}dz\int_{0}^{\infty}d\eta \delta\left(\sum_{i=1}^Nw_i-N\right)e^{-\beta\mathcal{H}\left(\{w_i\},z,\eta;\{x_{it}\}\right)}=\nonumber\\
=\int_{-\infty}^{\infty}\prod_{i=1}^Ndw_i\int_{-\infty}^{\infty}d\lambda e^{i\lambda\left(\sum_{i=1}^Nw_i-N\right)-\beta\frac{N}{T}\sum_{i=1}^N\sum_{t=1}^Tw_ix_{it}}\times\label{eq:partfunc}\\
\times\int_{0}^{\infty}dz\int_{0}^{\infty}d\eta e^{N\beta\phi\sqrt{z}+\beta\eta\left[\sum_{t=1}^T\left(\sum_{i=1}^N\left(x_{it}-\frac{1}{T}\sum_{s=1}^Tx_{is}\right)w_i\right)^2-Tz\right]}.\nonumber
\end{eqnarray}
Then the risk at the optimum, estimated from sample $\{x_{it}\}$, is computed as:
\begin{equation}
\hat{\mathcal{R}}_{\phi}^{*}=-\lim_{\beta\to\infty}\frac{1}{\beta N}\log Z_{\beta}\left[\phi;\{x_{it}\}\right].
\end{equation}
This is nothing but the free energy density at zero temperature (i.e. the ground state energy density). 

The free energy and all the "thermal averages" one can derive from it depend on the random sample. In general, one is interested in computing averages over the sample fluctuations (e.g. $\mathbb{E}q_0^2$), so we have to average the free energy over the random samples. To obtain $\mathbb{E}\hat{\mathcal{R}}_{\phi}^{*}$ we have to compute $\mathbb{E}\log Z_{\beta}\left[\phi;\{x_{it}\}\right]$. Averaging the logarithm of a random variable is a hard task. The replica method (see e.g. \cite{mezparvir}) was invented to circumvent this difficulty by the use of the identity
\begin{equation}
\log{Z}=\lim_{n\to0}\frac{Z^n-1}{n},
\end{equation}
and computing $\mathbb{E}Z^n$ for positive integer $n$, which is a relatively simple task. In order to be able to take the $n\to0$ limit, ultimately one has to analytically continue to real $n$. The name of the method derives from the fact that $Z^n$ is the partition function of a system that consists of $n$ identical copies (replicas) of the original problem. The Achilles heel of the method is the analytic continuation whose uniqueness usually cannot be guaranteed; we will justify its use ex post by the simulation results to be presented in the next section.

The sample elements $x_{it}$ are independent and identically distributed random variables, so assuming a variance of $1/N$ their 
joint probability distribution function is
\begin{equation}
p(\{x_{it}\})=\left(\frac{N}{2\pi}\right)^{NT/2}\exp\left(-\frac{N}{2}\sum_{i=1}^N\sum_{t=1}^Tx_{it}^2\right).\label{eq:jpdf}
\end{equation}
We can compute $\mathbb{E}Z^n$, by expressing $Z^n$ as the product of $n$ independent, identical integrals over the replicated variables $w_i^a$, $z^a$ and $\eta^a$ ($a=1,2,...,n$), and then taking its average with respect to the density function \eref{eq:jpdf}. After computing several Gaussian integrals we arrive at the expression
\begin{equation}\fl
\mathbb{E}Z_{\beta}^n\left[\phi\right]\propto\int_{-\infty}^{\infty}dQ^{ab}\int_{-i\infty}^{i\infty}d\hat{Q}^{ab}\int_{0}^{\infty}dz\int_{0}^{\infty}d\eta
e^{NG_{\beta}(\{Q^{ab}\},\{\hat{Q}^{ab}\},\{z^a\},\{\eta^a\})}\label{eq:avgpartfunc}
\end{equation}
where we omitted the normalizing factor and used the notations
\begin{eqnarray}
\fl G_{\beta}(\{Q^{ab}\},\{\hat{Q}^{ab}\},\{z^a\},\{\eta^a\})=\nonumber\\
=\sum_{a,b=1}^{n}\hat{Q}^{ab}\left(Q^{ab}-1\right)-\frac{1}{2}\Tr\log\hat{\mathbf{Q}}-\frac{1}{2r}\Tr\log\mathbf{Q}-\\
-\beta\sum_{a=1}^n\phi\sqrt{z^a}+\frac{\beta}{r}\sum_{a=1}^n\eta^az^a+\frac{1}{N}\log A_{\beta}(\{Q^{ab}\},\{\eta^a\}),\nonumber
\end{eqnarray}
and
\begin{eqnarray}
\fl A_{\beta}(\{Q^{ab}\},\{\eta^a\})=\int_{-\infty}^{\infty} du_t^a \exp\left[-\frac{1}{2}\sum_{a,b=1}^n\left(\mathbf{Q}^{-1}\right)^{ab}\sum_{t=1}^Tu_t^au_t^b+\beta r\sum_{a=1}^n\sum_{t=1}^Tu_t^a\right]\times\nonumber\\
\times\exp\left[-\beta\sum_{a=1}^n\sum_{t=1}^T\eta^a{u_t^a}^2+\beta\frac{r}{N}\sum_{a=1}^n\eta^a\left(\sum_{t=1}^T u_t^a\right)^2\right].
\end{eqnarray}
Here we introduced the so called overlap matrix
\begin{equation}
Q^{ab}=\frac{1}{N}\sum_{i=1}w^a_iw^b_i.
\end{equation}
and its conjugate $\hat{Q}^{ab}$, which is a Lagrange multiplier to enforce the equality above. As we are interested in the $N\to\infty$ limit, we can use the saddle point method to compute the integral \eref{eq:avgpartfunc}. Since we are dealing with a convex optimization problem, we expect that the saddle point is replica symmetric, that is we assume that $Q^{ab}=q+\Delta q\delta^{ab}$, $\hat{Q}^{ab}=\hat{q}+\Delta\hat{q}\delta^{ab}$, $\eta^a=\eta$ and $z^a=z$. After eliminating $\hat{q}$ and $\Delta\hat{q}$ by partial extremization, we get $G_{\beta}(q,\Delta q,z,\eta)=n[g_0+\beta g_{\beta}(q,\Delta q,z,\eta)]+\Or(n^2)$, where $g_0$ is some constant and 
\begin{eqnarray}
\fl g_{\beta}(q,\Delta q,z,\eta)=-\frac{1}{2\beta\Delta q}-\frac{1-r}{2\beta r}\left(\log\Delta q+\frac{q}{\Delta q}\right)-\nonumber\\
-\phi\sqrt{z}+\frac{1}{r}z\eta+\frac{1}{\beta Nn}\log A(q,\Delta q,\eta)\\
\fl A_{\beta}(q_0,\Delta q,\eta)=\int_{-\infty}^{\infty} du_t^a \exp\left[\frac{q}{2\Delta q^2}\sum_{t=1}^T\left(\sum_{a=1}^nu_t^a\right)^2-\frac{1}{2\Delta q}\sum_{a=1}^n\sum_{t=1}^T{u_t^a}^2\right]\times\nonumber\\
\times\exp\left[-\beta\eta\sum_{a=1}^n\sum_{t=1}^T{u_t^a}^2+\beta r\sum_{a,t}u_t^a+\beta\eta\frac{r}{N}\sum_{a=1}^n\left(\sum_{t=1}^T u_t^a\right)^2\right]
\end{eqnarray}
In the thermodynamic limit, the optimum can be obtained by minimizing the free energy density, which works out to be
\begin{equation}
f_{\beta}(q,\Delta q,z,\eta)=-\frac{1}{\beta}\lim_{N\to\infty}\frac{1}{N}\lim_{n\to 0}g_{\beta}(q,\Delta q,z,\eta).
\end{equation}
In this limit $\log A_{\beta}(q_0,\Delta q,\eta)/Nn$ can be computed explicitly by performing the Hubbard-Stratonovich transformation twice to linearize quadratic terms in the exponent of the integrand, then computing a few more Gaussian integrals and approximating the logarithm function by its series expansion around 1. Finally we get 
\begin{eqnarray}\fl
f_{\beta}(q,\Delta,z,\eta)=\frac{1}{2\Delta}+\frac{1-r}{2r}\left(\frac{1}{\beta}\log\frac{\Delta}{\beta}+\frac{q}{\Delta}\right)+\phi\sqrt{z}-\frac{1}{r}z\eta+\nonumber\\
+\frac{1}{2r}\left[\frac{1}{\beta}\log\left(\frac{2\pi\Delta}{1+2\eta\Delta}\right)+\frac{q}{\Delta+2\eta\Delta^2}+\Delta r^2\right]\label{eq:fed}
\end{eqnarray}
where we introduced the variable $\Delta=\beta\Delta q$. It is clear that in the zero temperature ($\beta\to\infty$) limit the free energy density is finite only if $\Delta$ remains a non-zero, finite constant. (In other words, the difference between the diagonal and non-diagonal elements of the replica matrix is proportional to $\beta^{-1}$, therefore, it vanishes in the zero temperature limit.)

\begin{figure}
\begin{center}
\includegraphics[keepaspectratio,width=6cm]{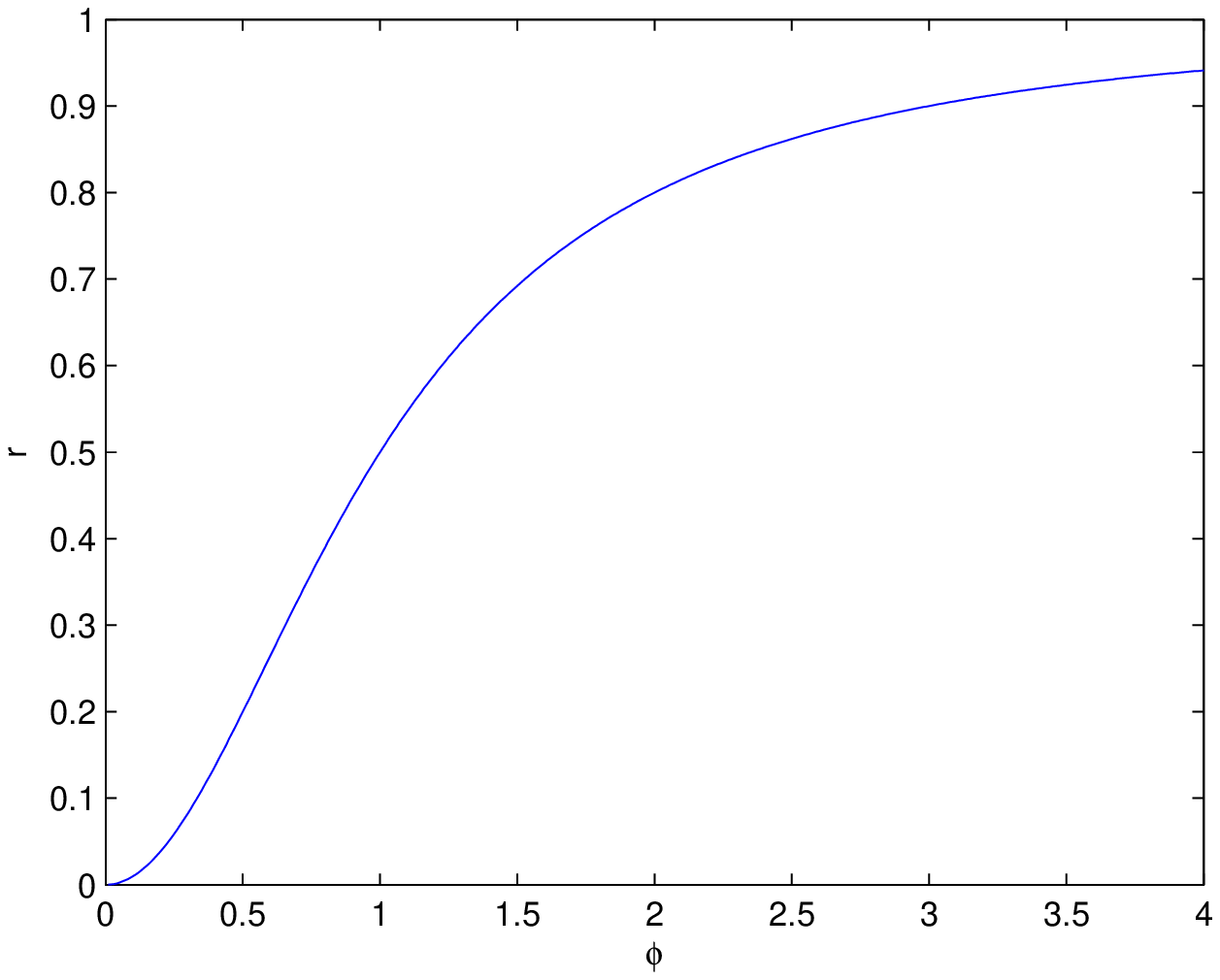}
\includegraphics[keepaspectratio,width=6cm]{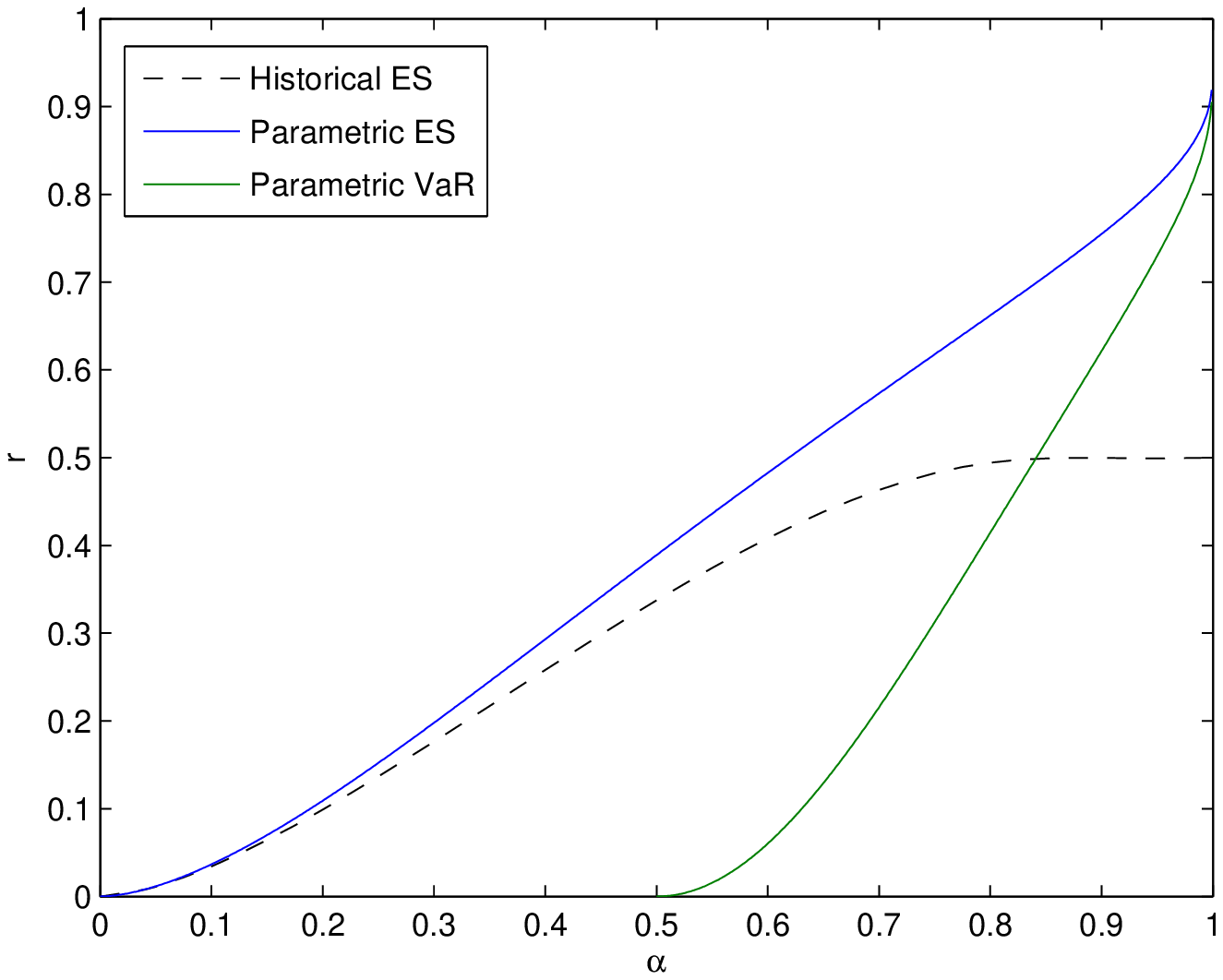}
\caption{\label{fig:theorphase} \textit{Left panel}: The curve of critical $N/T$ values as a function of $\phi$. \textit{Right panel}: The curves of critical $N/T$ values as a function of $\alpha$ for VaR and ES. For comparison purposes, the critical curve of the historical ES optimization problem is also plotted with a black dashed line.}
\end{center}
\end{figure}

Introducing the new variables $\eta'=2\eta\Delta$ and $q'=q/\Delta^2$ we obtain the zero temperature free energy density in the form:
\begin{equation}\fl
f_{0}(q',\Delta,z,\eta')=\phi\sqrt{z}-\frac{z\eta'}{2r\Delta}+\frac{1}{2\Delta}+\frac{\Delta}{2r}\left[\left(1+r+\frac{1}{1+\eta'}\right)q'+r^2\right].
\end{equation}
The saddle point conditions now read 
\begin{equation}
\frac{\partial f}{\partial q'}=\frac{\partial f}{\partial \Delta}=\frac{\partial f}{\partial z}=\frac{\partial f}{\partial \eta'}=0.
\end{equation}
which implies that the solution is
\begin{eqnarray}
q'^*=\phi^2\\
\Delta^*=\left[(1-r)\phi^2-r\right]^{-1/2}\label{eq:Delta}\\
\eta'^*=\frac{r}{1-r}\\
z^{*}=\frac{(1-r)^2}{4}\phi^2
\end{eqnarray}
From \eref{eq:Delta} it is clear that the saddle point method is only meaningful, if $(1-r)\phi^2-r>0$. That is, in the thermodynamic limit, for each value of $\phi$ there is a critical value $r_c$ of $r=N/T$ so that the optimization problem is not feasible unless $r<r_c$. (This stability condition corresponds to \eref{eq:condexist} in the thermodynamic limit.) \Eref{eq:Delta} implies that the critical values $r_c$ are on the curve
\begin{equation}
r_c(\phi)=\frac{\phi^2}{\phi^2+1},\label{eq:phasebound}
\end{equation}
which divides the $r$ vs $\phi$ plane into two distinct phases: one in which the optimization is feasible and another one in which it is not. The implied phase diagrams can be seen in \Fref{fig:theorphase}. The left panel shows the phase boundary in the $r$ vs $\phi$ plane.
It is interesting to take a look at the asymptotic behavior of $r_c(\phi)$: as it increases in a strictly monotonous manner and $\lim_{\phi\to\infty}r_c(\phi)=1$, it is clear that  $r_c(\phi)<1$ for any finite $\phi$. In other words, for any confidence level $\alpha<1$ (whether we are dealing with VaR or ES) the minimal length of the time series that ensures the existence of the optimum must be greater than $N$. 

Substituting the formulas in \eref{eq:phivares} into \eref{eq:phasebound} we get the phase boundaries of VaR and ES, respectively, in the $r$ vs $\alpha$ plane (right hand side of \Fref{fig:theorphase}). It can be seen that parametric VaR optimization is more unstable than the parametric optimization of ES, although for practically relevant values of $\alpha$ (that is in the $\alpha>.9$ range) the difference is not very significant. (For instance, for $\alpha=99\%$ the critical value $r_c$ is about $0.844$ and $0.877$ for VaR and ES respectively.) An interesting feature of both phase diagrams is that close to $\alpha=1$ they tend to $r=1$ with infinite derivatives.

The right panel of \Fref{fig:theorphase} also shows the phase boundary of historical ES, so we can easily compare it to the critical curve of parametric ES. It is clear that the non-parametric phase curve is below the parametric one for any confidence level $\alpha$, therefore the parametric estimation is more stable. In other words, a shorter time series is enough to ensure the feasibility of portfolio optimization, if parametric ES estimation is used. This was to be expected, but it is important to stress that although parametric fitting reduces the chance that there is no optimum for a given sample (especially for larger values of $\alpha$), it fails to completely eliminate the feasibility problem originally encountered in historical estimation \cite{kondor1}.

Let us now derive the sample average of the noise sensitivity measure $q_0^2$ in the thermodynamic limit, provided the optimum exists. Let us denote this conditional sample average by $\tilde{\mathbb{E}}$. From \eref{eq:q02def} and the replica symmetric ansatz it follows that $q_0^2=q+\Delta/\beta$. Therefore, in the $\beta\to\infty$ limit we find that the conditional average of the estimation error of the optimal portfolio is
\begin{equation}
\tilde{\mathbb{E}}q_0^2=q'^*\cdot{\Delta^*}^2=\frac{\phi^2(\alpha)}{(1-r)\phi^2(\alpha)-r}=\frac{r_c(\alpha)}{r_c(\alpha)-r}.\label{eq:q02}
\end{equation}
That is, $q_0\sim(r_c-r)^{-1/2}$, so the estimation error of the parametric VaR and ES optimization displays the same critical behavior as the minimization of variance, mean absolute deviation, maximal loss and non-parametric ES. More generally, it is very probable that the parametric ES and VaR belong to the same universality class as the aforementioned risk measures, which would imply that $q_0^2$ is self-averaging (that is its variance vanishes in the thermodynamic limit) also here. This is clearly supported by numerical simulations and should be possible to confirm by a (very hard) replica calculation which is, however, beyond the scope of this paper.

\subsection{Correlated asset returns}
\label{sec:corr}

In this section we are going to show that the results presented above will not change, if we allow asset returns to be correlated. We still assume that the data generating process is iid normal with zero expectations, but now we allow  the covariance matrix $\sigma_{ij}$ to be any strictly positive definite matrix. Therefore $\sigma_{ij}$ has a Cholesky-decomposition, in other words, there is a lower tirangular matrix $D_{ij}$ so that
\begin{equation}
\sigma_{ij}=\sum_{k=1}^ND_{ik}D_{jk}.
\end{equation}
Let $y_{it}$ ($i=1,2,...,N$, $t=1,2,...,T$) be a sample of normally distributed asset returns with zero mean and covariances $\sigma_{ij}$. It is easy to see that the variables
\begin{equation}
x_{it}=\sum_{j=1}^ND^{-1}_{ij}y_{jt}
\end{equation}
have a standard normal distribution, moreover, the observed return of a portfolio $\{v_i\}$ over the time period $t$ can be written as
\begin{equation}
\sum_{i=1}^Nv_iy_{it}=\sum_{i=1}^Nv_i\sum_{j=1}^ND_{ij}x_{it}=\sum_{i=1}^Nw_ix_{it},
\end{equation}
where we introduced the notation
\begin{equation}
w_i=\sum_{j=1}^Nv_jD_{ji}.
\end{equation}
Hence, the matrix $D_{ij}$ defines a linear transformation under which the scalar products between the asset return vectors and the portfolio vectors are invariant. This immediately implies that the Hamiltonian \eref{eq:hamilt} as well as $q_0^2$ are also invariant under this transformation, because they only depend on the observed asset returns and the portfolio vector through their scalar products. (It is important to bear in mind that this is only true, if the expected values of the asset returns are zero.) The budget constraint equation is not invariant, however, and it will take the form
\begin{equation}
\sum_{i=1}^Nw_i\sum_{j=1}^ND^{-1}_{ij}=N.\label{eq:corrbudget}
\end{equation}
The financial interpretation of this result is straightforward. For each $i$ the vector defined by $\{d^{(i)}_j\}_j=\{D^{-1}_{ij}\}_j$ can be regarded as a portfolio. Then $x_{it}$ denotes the return of  $\{d^{(i)}_j\}_j$ in the time interval $t$. So the vector $\{w_i\}$ is an equivalent representation of the portfolio $\{v_i\}$, but while the latter is expressed in terms of the original, correlated assets, the numbers $\{w_i\}$ specify the weights of the standard normal assets $\{d^{(i)}_j\}_j$. Since the vectors $\{d^{(i)}_j\}_j$ are not normalized in general (their components do not sum to unity), the weights $\{w_i\}$ are measured in different units than $\{v_i\}$. This is why the components $w_i$ have to be rescaled in the transformed budget constraint \eref{eq:corrbudget}.

As a result, the partition function for the sample $\{y_{it}\}$ of correlated asset returns can be expressed in terms of the standard normal variables $x_{it}$:
\begin{eqnarray}\fl
Z_{\beta}\left[\phi;\{x_{it}\}\right]=\int_{-\infty}^{\infty}\prod_{i=1}^Ndw_i\int_{-\infty}^{\infty}d\lambda e^{i\lambda\left(\sum_{i=1}^Nw_i\sum_{j=1}^ND^{-1}_{ij}-N\right)-\beta\frac{N}{T}\sum_{i=1}^N\sum_{t=1}^Tw_ix_{it}}\times\\
\times\int_{0}^{\infty}dz\int_{0}^{\infty}d\eta e^{N\beta\phi\sqrt{z}+\beta\eta\left[\sum_{t=1}^T\left(\sum_{i=1}^N\left(x_{it}-\frac{1}{T}\sum_{s=1}^Tx_{is}\right)w_i\right)^2-Tz\right]}.\nonumber
\end{eqnarray}
This expression is very similar to \eref{eq:partfunc} and the replica calculations presented in the previous section can be repeated to derive the quenched average of the free energy:
\begin{eqnarray}\fl
f_{\beta}(q,\Delta,z,\eta)=\frac{\gamma}{2\Delta}+\frac{1-r}{2r}\left(\frac{1}{\beta}\log\frac{\Delta}{\beta}+\frac{q}{\Delta}\right)+\phi\sqrt{z}-\frac{1}{r}z\eta+\nonumber\\
+\frac{1}{2r}\left[\frac{1}{\beta}\log\left(\frac{2\pi\Delta}{1+2\eta\Delta}\right)+\frac{q}{\Delta+2\eta\Delta^2}+\Delta r^2\right]\label{eq:fedcorr}
\end{eqnarray}
which only differs from \eref{eq:fed} in the first term $\gamma/2\Delta$ where
\begin{equation}
\gamma^{-1}=\lim_{N\to\infty}\frac{1}{N^2}\sum_{i=1}^N\sum_{j=1}^N\sigma^{-1}_{ij}.\label{eq:gamma}
\end{equation}
This means that for different values of $N$ the covariance matrix $\sigma_{ij}$ can be chosen freely, the only restriction is that the limit \eref{eq:gamma} must exists. Moreover, without restricting generality, the existence of this limit can always be ensured by fixing a positive number $\gamma$ at our convenience and rescaling the asset returns by a constant for each finite value of $N$ so that $N^2/\sum_{i,j}\sigma^{-1}_{ij}$ is equal to $\gamma$. Since $\gamma$ is arbitrary, it is clear that the phase boundary as well as $\tilde\mathbb{E}(q_0^2)$ must not depend on it. In fact, the minimization of the free energy density \eref{eq:fedcorr} yields exactly the same results as in the uncorrelated case, namely \eref{eq:phasebound} and \eref{eq:q02}.

Finally, it is worth noting that the free energy density of the uncorrelated problem can be recovered from \eref{eq:fedcorr} by letting $\sigma_{ij}=N^{-1}\delta_{ij}$, so that $\gamma=1$. The variance scaling factor $N^{-1}$ is in deed reflected in the joint probability density function \eref{eq:jpdf} used to average over the uncorrelated asset returns.

\subsection{Numerical study}
\label{sec:montecarlo}

In view of the heuristic character of the replica computation, we feel it is useful to provide numerival evidence to support its results. In order to do this, we generated  independent samples from a multivariate standard normal distribution ($\mu_i=0$ and $\sigma_{ij}=\delta_{ij}$), and attempted to find the minimum of $\hat{\mathcal{R}}_{\phi}(\{x_{it}\})$ in each sample. For the sake of simplicity, rather than controlling the value of $\alpha$, we controlled $\phi$ directly. To measure the probability of the existence of a minimum for a given combination of $N$, $T$ and $\phi$, we used the following algorithm:
\begin{enumerate}
\item \label{item:first} Generate an $N\times T$ sample matrix $\{x_{it}\}$.
\item Estimate the means and the covariances from $\{x_{it}\}$ using equations \eref{eq:meanest} and \eref{eq:covest}.
\item \label{item:last} Use the condition \eref{eq:condexist} to check if the portfolio optimization problem is feasible on the sample $\{x_{it}\}$.
\item Repeat steps (\ref{item:first}) to (\ref{item:last}) $K$ times, and count how many times the optimum exists. Let this number be $L$. Then the estimated probability of feasibility will be $\hat{p}(N,T,\phi)=L/K$.
\end{enumerate}
Clearly, the larger $K$ the more accurate the measurement will be.

\begin{figure}
\begin{center}
\includegraphics[keepaspectratio,width=6cm]{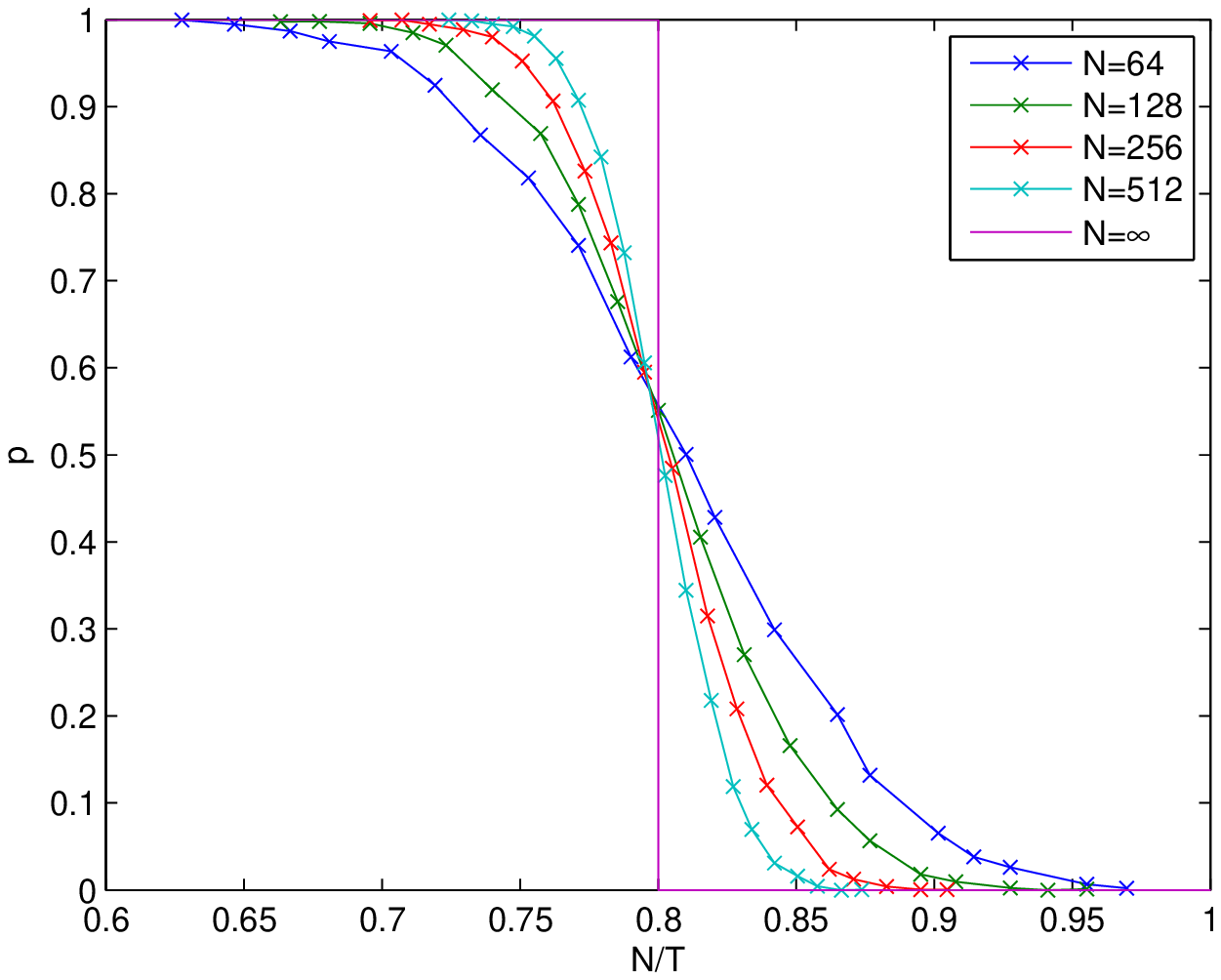}
\includegraphics[keepaspectratio,width=6cm]{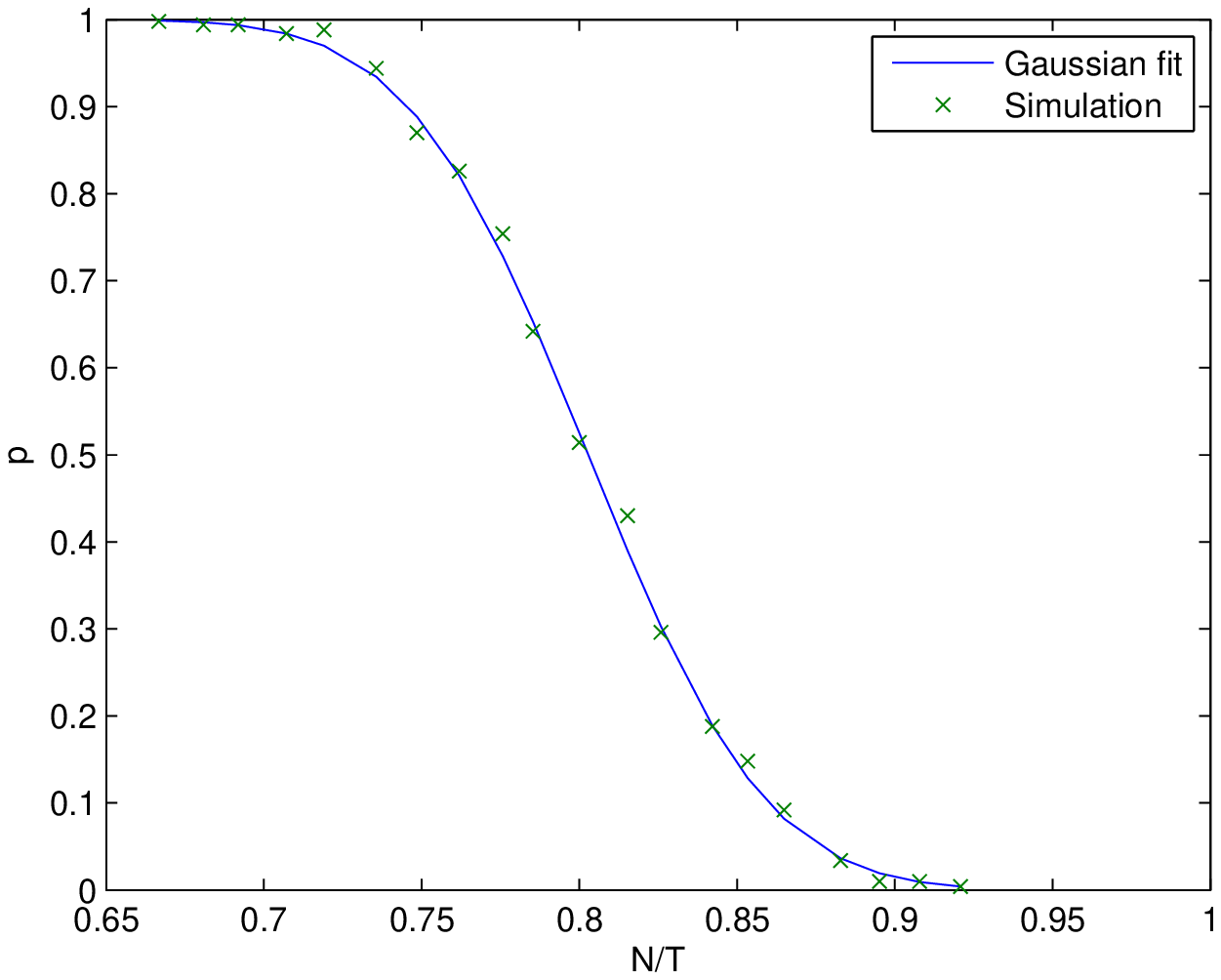}
\caption{\label{fig:p_vs_r} \textit{Left panel}: The estimated probability $p$ of the existence of an optimum as a function of $N/T$ for $\phi=2$ and different values of $N$. The curve labeled by $N=\infty$ corresponds to the thermodynamic limit (as computed by the replica method). \textit{Right panel}: Gaussian curve fitting to the measured probabilities for $N=128$ and $\phi=2$.}
\end{center}
\end{figure}

The left panel of \Fref{fig:p_vs_r} exhibits simulation results for $\phi=2$, which corresponds to confidence levels of $\alpha=0.9772$ for VaR and $\alpha=0.9420$ for ES. The number of iterations was $K=2000$ and the $p$ vs $\phi$ curve was measured for different values of $N$ (64, 128, 256 and 512). Based on the previous section, the critical value of $N/T$ is $r_c=0.8$, that is, in the thermodynamic limit the optimum exists with probability 1 if $N/T<0.8$ and it abruptly drops to 0 at the critical value (this is represented by the curve labeled by $N=\infty$ in the figure). The diagram shows that for finite values of $N$ and $T$ the probability of the existence of the optimal portfolio decreases from 1 to 0 continuously. At the same time, as $N$ increases (that is, as we approach the thermodynamic limit) the fall of the probability from 1 to 0 becomes sharper and sharper, as expected. The probability curves belonging to different values intersect one another at the same point, therefore, this point must correspond to the critical value $r_c$. As shown by the figure, the intersection is, indeed, very close to $r=0.8$, in excellent agreement with the analytical results.

We also observed that the probability curves fit very well to the function $g_{\mu,\sigma}(x)=1-\Phi((x-\mu)/\sigma)$ where $\Phi(x)$ is the cumulative distribution function of the standard normal distribution, and $\mu$ and $\sigma$ are parameters to be determined (e.g. via maximum likelihood estimation). The right hand panel shows simulated data points for $N=128$ and $\phi=2$ along with the fitted curve (where $\mu=0.8028$ and $\sigma=0.0446$). It is clear that $g_{\mu,\sigma}(x)$ cannot be the exact model for the $p$ vs $N/T$ curve, since for $N/T>1$ we have $p=0$. This fact, however, gradually loses its significance as $N$ increases, and $\sigma$ gets smaller and smaller. As a result, fitting $g_{\mu,\sigma}(x)$ to the numerically computed data points makes it possible to estimate $p$ as a function of $\phi$ and $N/T$ with a high accuracy, even if the number of iterations $K$ is low; this way simulations can be speeded up by a factor ranging from 10 to 100.

\begin{figure}
\begin{center}
\includegraphics[keepaspectratio,width=6cm]{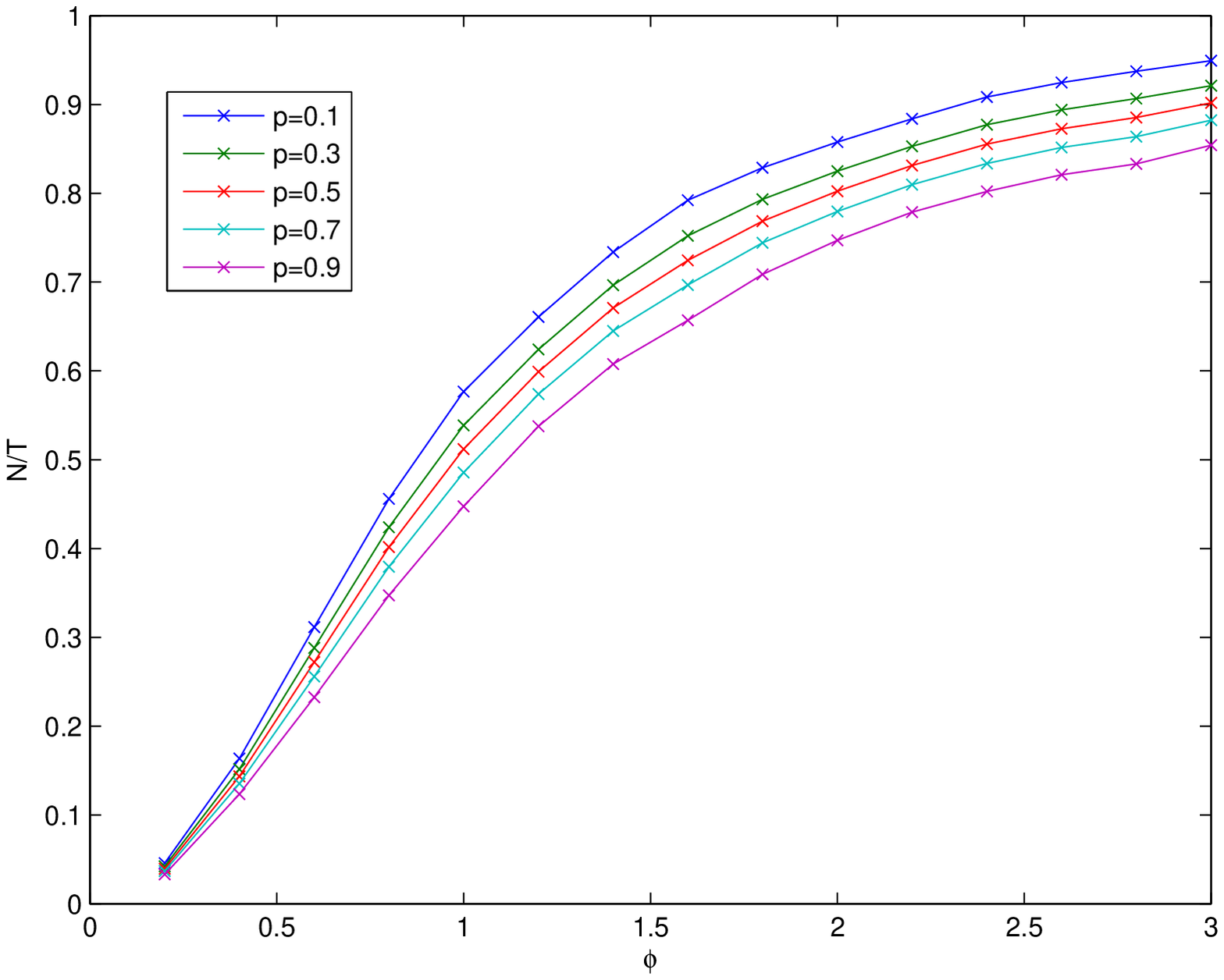}
\includegraphics[keepaspectratio,width=6cm]{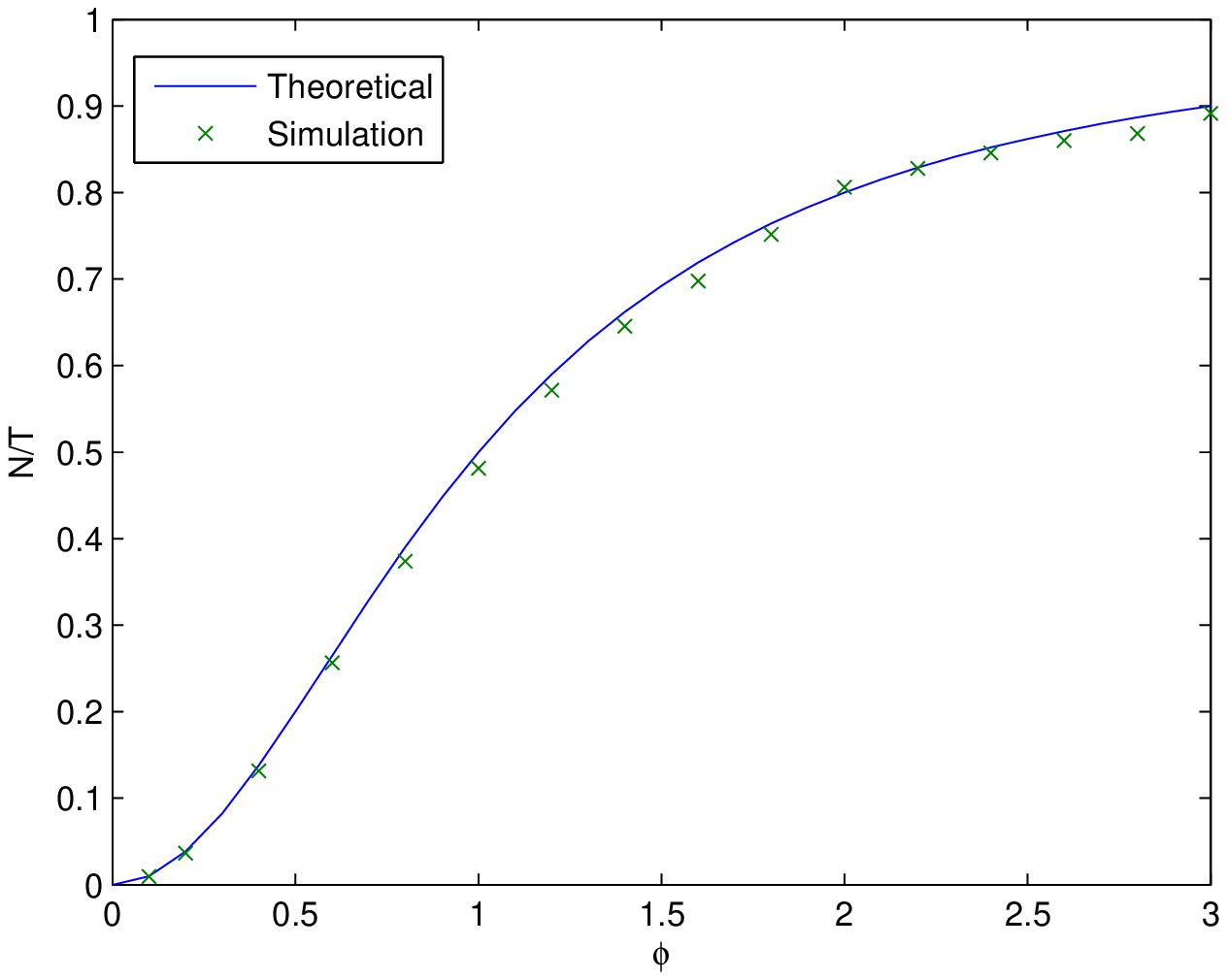}
\caption{\label{fig:phasediag} \textit{Left panel}: Contour lines of fixed $p$ for $N=128$, on the $N/T$ vs $\phi$ plane. \textit{Right panel}: Phase boundary in the $N\to\infty$ and $N/T$ finite limit.}
\end{center}
\end{figure}

Our numerical study showed that around the critical value $r_c(\phi)$ the probability $p(N,T,\phi)$ follows the behavior displayed in \Fref{fig:p_vs_r} for any value of $\phi$, but the steepness of the decline from 1 to 0 varies with $\phi$. To demonstrate this, we numerically computed the contour lines of constant $p$ on the $N/T$ vs $\phi$ plane for $p=0.1$, $0.3$, $0.5$, $0.7$ and $0.9$ with $N=128$ (the number of iterations was set to $K=100$, and we fitted $g_{\mu,\sigma}(x)$ to the simulated data points). The results are shown on the left hand side of \Fref{fig:phasediag}. Comparing this diagram to the left panel of \Fref{fig:theorphase} it is evident that the contour lines are arranged around, and have a similar shape to, the theoretical phase boundary. As mentioned above, the critical points can be estimated as the intersections of the $p$ vs $N/T$ curves for different values of $N$. The green points on the right hand panel were numerically computed by fitting $g_{\mu,\sigma}(x)$ to simulated data with $N=64$ and $N=128$, and then calculating the intersection of the two fitted curves (the number of iterations was $K=100$). The estimated critical points (in green) and the computed phase boundary (in blue) line up very well, which confirms the validity of the results obtained through the replica method.

\subsection{A note on semivariance}
\label{sec:semivar}

Semivariance is one of the oldest downside risk measures. As we shall see, the results obtained in the previous sections can be directly applied to characterize the stability of portfolio optimization under semivariance, when it is estimated by parametric fitting.

The definition of semivariance is
\begin{equation}
\nu^2(X)=\mathbb{E}\left[\max\{0,X-\mathbb{E}(X)\}\right]^2,
\end{equation}
where $X$ is a random variable representing the return of some security. The measure $\nu$ is simply called semi standard deviation, and this quantity can be used to define the following, VaR/ES-like risk measure (which is sometimesww called semivariance too, leading to some confusion):
\begin{equation}
\rho(X)=\nu(X)-\mathbb{E}(X).\label{eq:semivar}
\end{equation}
When the variable $X$ is normally distributed with mean $\mu$ and standard deviation $\sigma$ the semi standard deviation is simply $\nu=\sigma/\sqrt{2}$, so the risk measure $\rho$ can be written as
\begin{equation}
\rho(X)=\frac{1}{\sqrt{2}}\sigma-\mu,
\end{equation}
which is exactly of the same form as \eref{eq:paramrisk} with $\phi=1/\sqrt{2}\approx0.71$.

This implies immediately that in the case of semivariance minimization the critical value of $N/T$ is $r_c=1/3$, that is, for large $N$ (i.e. close to the thermodynamic limit), we need a time series that is at least three times as long as the number of assets in the portfolio, in order to have a meaningful optimization problem. Moreover, the conditional average of $q_0^2$ will be $\tilde{\mathbb{E}}q_0^2=(1/3-N/T)^{-1}/3$.

\section{Summary}
\label{sec:summary}

We studied the feasibility and noise sensitivity of portfolio optimization under Value-at-Risk, Expected Shortfall and semivariance in the case when these risk measures are estimated from finite samples using parametric fitting. Similarly to earlier studies based on non-parametric estimation \cite{kondor1,ciliberti1} we first assumed independent standard normal asset returns, but in our present work we generalized our results for correlated returns as well. We found that the probability that the optimum exists on a given finite sample is smaller than unity, and this probability is a function of the portfolio size, the sample size and the confidence level of VaR/ES. In the thermodynamic limit (where the portfolio size $N$ goes to infinity but its ratio to the sample size $T$ is held constant), this probability converges to either 0 or 1 depending on $N/T$ and the confidence level $\alpha$. We employed the replica method to compute the equation of the curve separating the feasible and unfeasible regions on the $N/T$ vs $\alpha$ plane, and also tested and supported the result by numerical simulation. The replica approach also enabled us to compute the average of the measure of noise sensitivity $q_0^2$, contingent on the feasibility of the optimization problem. It is highly probable that the parametric ES, VaR and semivariance optimization problems belong to the same universality class as the optimization of many other risk measures (standard deviation, mean absolute deviation, maximal loss, non-parametric ES): we found that the estimation error blows up with a critical exponent $-1/2$ as we approach the phase boundary.

Our results make it possible to compare the parametric and historical estimator of ES. It is clear that parametric estimation does not eliminate the instability of the historical estimator, but it does improve on it, in that the phase diagram of parametric ES runs above the historical curve. This means that for a given confidence level and a given portfolio size we need more data (longer time series) in the historical estimation than in the parametric one, in order to have a meaningful solution to the optimization problem. It seems as if we had some additional source of information in the parametric case. (The effect is even more pronounced in the case of VaR, where the historical estimate cannot be guaranteed to be convex for any confidence level and any length of the time series, whereas the parametric estimate has been shown here to have an optimum at least in a certain region of parameter space.) One may wonder where this additional information may have come from. The answer is simple: in the historical estimation we do not make any assumption about the nature of the underlying distribution, we are just using raw data as they are produced by the data generating process. In contrast, in the parametric case we assume that the process is Gaussian and fit the data to this assumption. This way we are projecting a nontrivial piece of information into the estimation. For technical reasons we have indeed chosen a Gaussian underlying process in the context of this work, but in a real market return fluctuations are neither Gaussian, nor even stationary. To project an arbitrary distribution into real, parsimonious data may produce apparently more stable estimates, but the gain may well turn out to be completely illusory and the results misleading. 

We would also like to draw attention to the fact that the critical value of the $N/T$ ratio depends on the risk measure and on the (historical or parametric) method of estimation. This critical ratio is never larger than $1$, and, depending on the risk measure and on the confidence level, it may be significantly smaller; e.g., as we have just seen, for the semivariance e.g. it is as low as $1/3$. This means that, depending on the risk measure, we need time series longer than two or three times the size of the portfolio, in order to have a solution at all, and much longer, in order to have a reliable estimate. In the context of portfolio selection, where, by the very nature of the task, the sampling frequency cannot be higher than once a week or even once a month, this condition is not easy to satisfy. Therefore, in practice the typical $N/T$ ratio may be fairly close to the phase boundary where the estimation error diverges. The knowledge of the phase boundary and the position of our working point (confidence level and $N/T$ ratio) relative to it is highly important if we wish to take sample to sample fluctuations properly into account. 

This work has presented further evidence for the instability of widely used risk measures against sample fluctuations. The instability of parametric VaR, easily the most popular risk estimate, is particularly notable. We find it remarkable how powerful the concepts and methods imported from the statistical physics of random systems prove to be in the analysis of these important phenomena.

\ack

This work has been supported by the ''Cooperative Center for Communication Networks Data Analysis'', a NAP project sponsored by the National Office of Research and Technology.

\section*{References}

\bibliography{varstability}
\bibliographystyle{unsrt}

\end{document}